\documentclass[letterpaper, 10 pt, conference]{ieeeconf}  % Comment this line out if you need a4paper

\IEEEoverridecommandlockouts                              % This command is only needed if 
\usepackage{amsfonts}
\usepackage{graphics} % for pdf, bitmapped graphics files
\usepackage{epsfig} % for postscript graphics files
\usepackage{mathptmx} % assumes new font selection scheme installed
\usepackage{times} % assumes new font selection scheme installed
\usepackage{amsmath} % assumes amsmath package installed
\usepackage{amssymb}  % assumes amsmath package installed
\usepackage{physics}
\usepackage{mdwmath}
\usepackage{mdwtab}
\usepackage[hidelinks]{hyperref}
\title{\LARGE \bf An Application of Pontryagin Neural Networks to Solve Optimal Quantum Control Problems}

\author{Nahid Binandeh Dehaghani and A. Pedro Aguiar % <-this % stops a space
\thanks{Research Center for Systems and Technologies (SYSTEC), Electrical and Computer Engineering Department, FEUP - Faculty of Engineering, University of Porto, Rua Dr. Roberto Frias sn, i219, 4200-465 Porto, Portugal
        {\tt\small nahid@fe.up.pt}
        {\tt\small pedro.aguiar@fe.up.pt}}% 
        }%

\begin{document}

\maketitle
\thispagestyle{empty}
\pagestyle{empty}

%%%%%%%%%%%%%%%%%%%%%%%%%%%%%%%%%%%%%%%%%%%%%%%%%%%%%%%%%%%%%%%%%%%%%%%%%%%%%%%%
\begin{abstract}
 Reliable high-fidelity quantum state transformation has always been considered as an inseparable part of quantum information processing. In this regard, Pontryagin maximum principle has proved to play an important role to achieve the maximum fidelity in an optimum time or energy. Motivated by this, in this work, we formulate a control constrained optimal control problem where we aim to  minimize time and also energy subjected to a quantum system satisfying the bilinear Schrödinger equation. We derive the first order optimality conditions through the application of Pontryagin Maximum (minimum) Principle, which results in a boundary
value problem. Next, in order to obtain efficient numerical results, we exploit a particular family of physics-informed neural networks that are specifically designed to tackle the indirect method based on the Maximum Principle of Pontryagin. This method has not yet been studied in the quantum context, but it can significantly speed up the process. To this end, we first obtain a set of relations which finally let us compute the optimal control strategy to determine the time- and energy-optimal protocol driving a general initial state to a target state by a quantum Hamiltonian with bounded control. We make use of the so-called "qutip" package in python, and the newly developed "tfc" python package. 

\end{abstract}

%%%%%%%%%%%%%%%%%%%%%%%%
%%%%%%%%%%%%%%%%%%%%%%%%%%%%%%%%%%%%%%%%%%%%%%%%%%%%%%%%%%%%%%%%%%%%%%%%%%%%%%%%
\section{INTRODUCTION}
 Over the last years, understanding the optimization methods of quantum systems has been a challenge for researchers. In the most quantum control protocols, the control law is preferably computed without experimental feedback in an open-loop layout. In this regard, Optimal Control Theory (OCT) provides powerful tools, allowing to formulate quantum control problems in order to seek a set of admissible controls satisfying the system dynamics while optimizing a cost functional (e.g., fidelity, time or energy). For optimizing a cost functional in an optimal control setup, several numerical solution methods have been developed, which can generally be divided into direct and indirect subcategories, \cite{graichen2010handling}. In the former method, the optimal control problem is considered as a non-linear optimization/programming problem, while the latter method is based on the optimality conditions of Pontryagin Maximum (or Minimum) Principle (PMP), which results in a two-point boundary value problem (BVP) in the state and adjoint conditions that possibly can be solved by shooting methods. As an example, a time discretized computational scheme has been proposed in \cite{dehaghani2022high}, aiming to solve a high fidelity quantum state transfer problem by means of an indirect method based on PMP.
 
Generally, dynamical quantum control originates from a time dependent quantum-mechanical Hamiltonian that steers the dynamics of the quantum system subjected to the constraints, e.g., relaxation processes. Hence, the realization of an optimal sequence is highly complicated in quantum control problems, and is dependant on the particular system under study. One way to deal with the problem complexity is the introduction of physics-informed neural networks (PINNs) to the context of quantum control. The term physics-informed neural networks is used to indicate the neural networks (NNs) for which the loss function comprises the physics as a regularization term, \cite{raissi2019physics}. PINNs can be applied to any quantum evolution with a well-known model. Recently, the flexibility of PINNs has been shown in quantum control context, \cite{norambuena2022physics}. 

In this work, we implement a quantum optimal control problem by applying the PMP and modeling a NN representation of the state-adjoint pair, for which the residuals of the BVP are considered as the loss to be as close to zero as possible. We will utilize the PINN methodology, developed in \cite{d2021pontryagin,johnston2021theory}, in quantum framework. In this method, PINNs are trained to learn actions satisfying PMP, so the optimality is guaranteed by learning the solutions of the associated BVP. These PINNs, named Pontryagin NNs (PoNNs), play an effective role in learning optimal control actions. Finally, considering the importance of quantum state to state transition problem in quantum information processing, we propose a time- and energy-minimum optimal control problem to achieve a high fidelity quantum state transfer through exploiting PoNNs.

The paper is organized as follows: We first recall the set of PMP optimality conditions for a general unconstrained quantum optimal control problem. Then, we review how PoNNs can be employed to learn the optimal control actions from the unknown solutions of the BVP arising from the PMP necessary optimality conditions. Afterwards, we formulate a quantum state transition problem under control constraints and obtain the necessary conditions of optimality. We then use PoNNs to solve the BVP, which is resulted from the application of PMP through the indirect method. The simulation results have been shown for the quantum state transfer problem in a three-level system. The paper ends with conclusions and an overview on prospective research challenges. 

\paragraph*{Notation.} For a general continuous-time trajectory $x$, the term $x(t)$ indicates the trajectory assessed at a specific time $t$. For writing the transpose of a matrix (or vector) we use the superscript $T$, and we use $\dagger$ to show the conjugate transpose of a matrix (or vector). Throughout the text, the vectors are shown by vectorbold command. To denote the wave functions as vectors, we use the Dirac notation such that $\left| \psi  \right\rangle =\sum\limits_{k=1}^{n}{{{\alpha }_{k}}\left| {{{\hat{\psi }}}_{k}} \right\rangle }$, where $\vert \psi \rangle $ indicates a state vector, ${\alpha }_{k}$ are the complex-valued expansion coefficients, and $\vert {\hat \psi}_k \rangle $ are basis vectors that are fixed. The notation bra is defined such that $\left \langle \psi  \right |= \left| \psi  \right\rangle^\dagger$. For writing partial differential equations (PDEs), we denote partial derivatives using subscripts. In the general situation where $f$ denotes a function of $n$ variables including $x$, then $f_x$ denotes the partial derivative relative to the $x$ input.

\section{An Overview on Quantum Optimal Control}

In quantum mechanics, a pure state is described by a unit vector wave function $\left| \psi \right\rangle$ in a complex Hilbert space $\mathbb{H}$, represented as 
\begin{equation}\label{1}
\left| \psi  \right\rangle =\cos \frac{\theta }{2}\left| 0 \right\rangle +{{e}^{i\varphi }}\sin \frac{\theta }{2}\left| 1 \right\rangle
\end{equation}
in which $\theta \in \left[ 0,\pi  \right]$, and $\varphi \in \left[ 0,2\pi  \right]$. 
The evolution of an n-level pure state $\left| \psi(t)  \right\rangle$ can be modelled through the bilinear time-dependent Schrödinger equation for closed quantum systems as 
\begin{equation}\label{2}
i\hbar \frac{\partial }{\partial t}\left| \psi \left( t \right) \right\rangle = H\left( t \right)\left| \psi \left( t \right) \right\rangle
\end{equation}
%, \quad \left| \psi (t) \right\rangle_{t=0} =\left| {{\psi }_{0}} \right\rangle
where $\left| \psi(t)  \right\rangle$ takes values on the state-space $S^{2n-1}$, and $\hbar$ is the reduced Planck’s constant, considered as a unit in the rest of the paper. The control evolution is determined through the quantum-mechanical system Hamiltonian ${H(t)}$, represented by
\begin{equation}\label{3}
\begin{aligned}
H\left( u(t) \right)&={{H}_{d}}+{{H}_{C}}\left( t \right)\\
&={{H}_{d}}+\sum\limits_{k=1}^m{{{u}_{k}}(t)}{{H}_{k}} \\
\end{aligned}
\end{equation}
The first term indicates the Hermitian matrix known as a drift Hamiltonian $H_d=diag(E_1,E_2,\cdots,E_n)$, where $E_k$ represents a real number concerning the energy level. The second term captures a set of external control functions $u_k(t)\in \mathbb{R}$ coupled to the quantum system via time independent interaction Hamiltonians ${{H}_{k}}$, which are Hermitian matrices to describe the coupling between the control and the system. One important objective of optimal control in quantum information processing is to the address state to state transition problem in the state space, so before going ahead, we must address the problem of controllability to check if it is possible to steer the system from one state to another for all pairs of possible states. To do so, we first check the controllability of the system evolution operator $U(t)$ satisfying
\begin{equation}\label{a}
i\frac{\partial }{\partial t}U \left( t \right)  = H\left( t \right)U(t)
\end{equation}
which is right-invariant on the compact Lie group $su(n)$. Note that it satisfies the necessary and sufficient condition of controllability since
\begin{equation}\label{b}
Lie\{iH_d,iH_1,\cdots,iH_m\}=su(n)
\end{equation}
which results in the controllability of \eqref{2} simply because the controllability of a right invariant system results in the controllability of the bilinear system. 

We now proceed with the optimal control formulation. For convenience and in order to express the problem in terms of only real quantities, we implement the system state and Hamiltonian as, \cite{d2021introduction, dehaghani2022optimal},
\begin{equation}\label{4}
\left| \Tilde{\psi} \left( t \right) \right\rangle =\left[ \begin{matrix}
   \operatorname{Re}\left( \left| \psi \left( t \right) \right\rangle  \right)  \\
   \operatorname{Im}\left( \left| \psi \left( t \right) \right\rangle  \right)  \\
\end{matrix} \right]
\end{equation}
and 
\begin{equation}\label{5}
\tilde{H}\left( u\left( t \right) \right):=\left( \begin{matrix}
   \operatorname{Re}\left( -iH\left( u\left( t \right) \right) \right) & -\operatorname{Im}\left( -iH\left( u\left( t \right) \right) \right)  \\
   \operatorname{Im}\left( -iH\left( u\left( t \right) \right) \right) & \operatorname{Re}\left( -iH\left( u\left( t \right) \right) \right)  \\
\end{matrix} \right)
\end{equation}
so we can rewrite \eqref{2} as 
\begin{equation}\label{6}
    \frac{\partial }{\partial t}\left| \tilde{\psi }\left( t \right) \right\rangle =\tilde{H}\left( u\left( t \right) \right)\left| \tilde{\psi }\left( t \right) \right\rangle 
\end{equation}

In the following, we deal with a problem to find a way to drive an initial state $\left|  \Tilde{\psi} \left( {{t}_{0}} \right) \right\rangle =\left| {{\psi }_{0}} \right\rangle$ to a desired final state $\left|  \Tilde{\psi} \left( {{t}_{f}} \right) \right\rangle =\left| {{\psi }_{f}} \right\rangle$ while minimizing a cost functional. The methods to design the quantum optimal controller vary according to the choice of the cost functional, the construction of the Pontryagin-Hamiltonian, and the computation scheme using the Maximum Principle conditions. Let us consider the following unconstrained optimal control problem ($OCP_u$), 
\begin{equation}\label{7}
\begin{aligned}
 & \textit{minimize}\quad  J\left( \left|  \Tilde{\psi} \left( t \right) \right\rangle ,u\left( t \right),t \right) \\ 
 & \quad =\Phi \left( \left| {{\psi }_{0}} \right\rangle ,\left( {{t}_{0}} \right),\left| {{\psi }_{f}} \right\rangle ,\left( {{t}_{f}} \right) \right)+\int\limits_{{{t}_{0}}}^{{{t}_{f}}}{L\left( \left|  \Tilde{\psi} \left( t \right) \right\rangle ,u\left( t \right),t \right)} dt \\ 
 & \textit{subjected to}  \\ 
 & \left| \dot{\Tilde{\psi}} \left( t \right) \right\rangle =f\left( \left| \Tilde{\psi} \left( {{t}} \right) \right\rangle,u\left( t \right),t \right) \\ 
 & \Phi \left( \left| {{\psi }_{0}} \right\rangle,{{t}_{0}} \right)={{\Phi }_{0}} \\ 
 & \Phi \left(\left| {{\psi }_{f}} \right\rangle,{{t}_{f}} \right)={{\Phi }_{f}} \\ 
\end{aligned}
\end{equation}
where $J$ in the cost function in which the term $\Phi$ indicates the end-point cost, and $L$ is the Lagrangian.
The state and control are indicated by $\left| \Tilde{\psi} \left( {t} \right) \right\rangle$ and $u(t)$, respectively, and $t$ in the independent variable with $t_0$ showing the initial and $t_f$ as the final time. For applying PMP within the indirect method, the Pontryagin Hamiltonian is defined by adjoining the constraints on the system dynamics to the Lagrangian ${\displaystyle L}$ via introducing time-varying Lagrange multiplier vector $\left| \lambda \left( {t} \right) \right\rangle$, whose elements are called the costates of the system. Hence, the Pontryagin Hamiltonian $\mathcal{H}$ for all ${\displaystyle t\in [t_0,t_f]}$ is constructed as
\begin{equation}\label{8}
\begin{aligned}
&\! \! \! \!  \!  \! \! \!  \! \! \! \! \!  \! \! \!  \mathcal{H}(\left| \tilde{\psi }\left( t \right) \right\rangle,u(t),\lambda (t),t) \\ 
&\quad = \left| \lambda \left( {t} \right) \right\rangle ^{{T}}\tilde{H}\left( u\left( t \right) \right)\left| \tilde{\psi }\left( t \right) \right\rangle +L(\left| \tilde{\psi }\left( t \right) \right\rangle),u(t))\\ \end{aligned}
\end{equation}
The first-order necessary optimality conditions of the PMP are then derived as
\begin{equation}\label{9}
    \frac{\partial \mathcal{H}}{\partial u}=0 
\end{equation}
\begin{equation}\label{10}
    {\left| \dot { \tilde{\psi }} \right\rangle}=\frac{\partial \mathcal{H}}{\partial \left| \lambda  \right\rangle}
\end{equation}
\begin{equation}\label{11}
    \dot{ \left| \lambda  \right\rangle }=-\frac{\partial \mathcal{H}}{\partial \left| \tilde{\psi } \right\rangle}
\end{equation}
Equations \eqref{10} and \eqref{11} that represent a boundary value problem are accompanied by the following possible transversality conditions on the Hamiltonian 
\begin{equation}\label{12}
   \left| \lambda \left( {{t}_{0}} \right) \right\rangle =-\frac{\partial J}{\partial \left| {{\psi }_{0}} \right\rangle}
\end{equation}
\begin{equation}\label{13}
   \mathcal{H} \left( {{t}_{0}} \right)=\frac{\partial J}{\partial {{t}_{0}}}
\end{equation}
\begin{equation}\label{14}
   \left| \lambda \left( {{t}_{f}} \right) \right\rangle =\frac{\partial J}{\partial \left| {{\psi }_{f}} \right\rangle}
\end{equation}
\begin{equation}\label{15}
   \mathcal{H} \left( {{t}_{f}} \right)=-\frac{\partial J}{\partial {{t}_{f}}}
\end{equation}
represent a boundary value problem (BVP). 

\section{Physics-Informed Neural Networks Based on the Theory of Functional Connections}
The boundary value problem resulted from the necessary optimality conditions has already been solved with several methods such as the so-called shooting method for quantum optimal control problems. In this section, we gave an overview of the newly-developed Pontryagin neural network method in \cite{d2021pontryagin}, and, then, we apply the formulations in quantum context. The main feature of PINNs is to embed the differential equations (DEs) and boundary conditions describing the physics of the problem in their cost functions.
PINNs can be designed to solve two classes of problems, including data-driven solution and data-driven discovery of PDEs, \cite{raissi2019physics}, following the original idea of using artificial neural networks to solve ordinary differential equations (ODEs) and PDEs in \cite{lagaris1998artificial}. PINNs are a kind of universal function approximators that by means of a single neural network can approximate the solution of differential equations. To better understand the procedure of PINNs, let consider the following differential equation of a generic BVP:
\begin{equation}\label{22}
\begin{aligned}
  & {{F}_{i}}\left( t,{{y}_{j}}\left( t \right),{{{\dot{y}}}_{j}}\left( t \right),{{{\ddot{y}}}_{j}}\left( t \right) \right)=0 \\ 
 & \textit{subjected to }:\left\{ \begin{matrix}
   {{y}_{j}}\left( {{t}_{0}} \right)={{y}_{{{0}_{j}}}}  \\
   {{y}_{j}}\left( {{t}_{f}} \right)={{y}_{{{f}_{j}}}}  \\
   {{{\dot{y}}}_{j}}\left( {{t}_{0}} \right)={{{\dot{y}}}_{{{0}_{j}}}}  \\
   {{{\dot{y}}}_{j}}\left( {{t}_{f}} \right)={{{\dot{y}}}_{{{f}_{j}}}}  \\
\end{matrix} \right. \\ 
\end{aligned}
\end{equation}
in which $i$ indicates the number of DEs forming the ODE system, and $j$ shows the number of unknown solutions of the system that are approximated by a NN. By considering ${\theta}$ as the NN parameters, which are trained by means of gradient-based methods, the PoNN is represented by, \cite{d2021pontryagin}, 
\begin{equation}\label{23}
{{y}_{j}}\left( t \right)=y_{j}^{NN}\left( t,\theta  \right)
\end{equation}
where the time $t$ is the input. PoNN is an especially trained NN satisfying the PMP through learning the solution of the arising BVP. To solve the BVP by applying the theory of functional connection (tfc) method, \cite{d2021pontryagin, johnston2021theory}, we first need to derive the constrained formulations together with their derivatives as
\begin{equation}\label{24}
    y_{j}^{m}\left( t,g\left( t \right) \right)=g_{j}^{m}\left( t \right)+\sum\limits_{k=1}^{{{n}_{j}}}{s_{k}^{m}\left( t \right){{\eta }_{kj}}}
\end{equation}
in which $g\left( t \right): \mathbb{R}\to \mathbb{R}$ indicates a user-specified function, $s_k\left( t \right): \mathbb{R}\to \mathbb{R}$ are the so-called support functions, and the coefficients $\eta_k$ represent constraint information, for all $n_j$ constraints. Here, the superscript $m$ and the subscript $j$ refer to the $m$th derivative and $j$th unknown function, respectively. The support functions are linearly independent, and can possibly be chosen as $s_k(t)=t^{k-1}$. By computing ${\eta }_{kj}$s, we can insert the boundary conditions to the constrained expression. Hence, by substituting the constrained expressions to \eqref{22}, we can obtain a new unconstrained set of equations, which are only a function of $g\left( t \right)$, represented as
\begin{equation}\label{25}
{{\tilde{F}}_{i}}\left( t,{{g}_{j}}\left( t \right),{{{\dot{g}}}_{j}}\left( t \right),{{{\ddot{g}}}_{j}}\left( t \right) \right)
\end{equation}
The free function ${{g}_{j}}\left( t \right)$, as developed in \cite{huang2006extreme} based on the theory of extreme learning machine (ELM), is modeled by a single hidden layer feedforward neural network (SLFN) as 
\begin{equation}\label{26}
    {{g}_{j}}\left( z \right)=\sum\limits_{n=1}^{L}{{{\xi }_{j,l}}h\left( {{\omega }_{l}}z_j+{{b}_{l}} \right)}=\xi _{j}^{T}{{h }_{j}}\left( z \right)
\end{equation}
where the summation is over all $L$ hidden neurons, and $h(\cdot)$ is the activation function. The output weight $\xi_l$ and the input weights $\omega_l=[\omega_{1},\omega_{2},\cdots,\omega_{L}]$ connect the $l$th hidden node to the output node and input nodes, respectively. $b_l$ indicates the threshold of the $l$th hidden node. Since the domain of $h(\cdot)$ and the $t$ domain are not synchronous, we must do a mapping between $t$ and $z$, as $z_j={{z}_{0}}+c\left( t_j-{{t}_{0}} \right)$, where $c>0$ is the mapping coefficient. Therefore, the derivatives of ${{g}_{j}}\left( t \right)$  are transformed from $t$ to $z$ domain as 
\begin{equation}\label{27}
    \frac{{{d}^{n}}{{g}_{j}}}{d{{t}^{n}}}=\xi _{j}^{T}\frac{{{d}^{n}}{{h}_{j}}\left( z \right)}{d{{z}^{n}}}{{c}^{n}}
\end{equation}
Since the theory of ELM is used for training the neural network, the unknowns would be ${{\xi}_{j}}={{\left[ {{\xi}_{j,1}},...,{{\xi}_{j,L}} \right]}^{T}}$. Moreover, for the free final time optimal control problems, the coefficient $c$ also becomes an unknown that needs to be computed. Now, we can summarize \eqref{25} as
\begin{equation}\label{28}
    {{\tilde{F}}_{i}}\left( {{z}_{j}},{{\xi }_{j}} \right)=0
\end{equation}
for which we need to discretize $z_j$ into $N$ points in order to obtain the set of unconstrained differential equations expressed as loss functions, so
\begin{equation}\label{29}
{{\mathcal{L}_{i}}\left( {{\xi }_{j}} \right)=\left\{ \begin{matrix}
\begin{aligned}
&{{{\tilde{F}}}_{i}}\left( {{z}_{0}},{{\xi}_{j}} \right)  \\
& {{{\tilde{F}}}_{i}}\left( {{z}_{1}},{{\xi}_{j}} \right) \\ 
&\vdots\\
& {{{\tilde{F}}}_{i}}\left( {{z}_{N-1}},{{\xi}_{j}} \right) \\ 
&{{{\tilde{F}}}_{i}}\left( {{z}_{N}},{{\xi}_{j}} \right)  \\
\end{aligned}  \\
\end{matrix} \right\}}
\end{equation}
For $\tilde{N}$ differential equations, \eqref{29} is augmented to the loss function $\mathbb{L}^{T}=\left[ {{\mathcal{L}
}_{1}},{{\mathcal{L}}_{2}},\ldots {{\mathcal{L}}_{i}},\ldots ,{{\mathcal{L}}_{{\tilde{N}}}} \right]$. By imposing a true solution, we have 
\begin{equation}\label{30}
\mathbb{L}=0_{N*\tilde{N}}
\end{equation}
Therefore, the ${{\xi}_{j}}$ coefficients can be learnt. It can be done by means of various optimization techniques, \cite{d2021pontryagin}.
\section{Quantum State Transition by a Pontryagin Neural Network Approach}
For many quantum operations, it is crucial to achieve the quantum state transition in the shortest possible time. However, for a given control objective, control amplitude is inversely proportional to the control time. Hence, in the case that state transition needs to be done in the shortest possible time, the control amplitude may be very large, which is not practically possible since the amplitude of control cannot be infinite. To address this problem, we now extend the optimal time-energy control problem formulation for the input constrained case. More precisely, in the sequel, we are interested in solving the input constrained optimal control problem ($OCP_c$):
\begin{equation*}
\begin{aligned}
    \min_{u,t_f} & & \left\{J = \Gamma\, {{t}_{f}}+\eta\int_{{{t}_{0}}}^{t_f}{{{u}^{2}}\left( t \right)\,dt}\right\} \\
    \mbox{subject to } &&\left| \dot{\Tilde{\psi}} \left( t \right) \right\rangle =f\left( \left| \Tilde{\psi} \left( {{t}} \right) \right\rangle,u\left( t \right),t \right)=\tilde{H}\left( u\left( t \right) \right)\left| \tilde{\psi }\left( t \right) \right\rangle\\
     && \left|  \Tilde{\psi} \left( {{t}_{0}} \right) \right\rangle =\left| {{\psi }_{0}}\right\rangle\in {{\mathbb{R}}^{n}}\\
     && {{t}_{0}}\le t\le {{t}_{f}} \\
    % && {{u}_{\min }}\le {{u}_{x,y}}\left( t \right)\le {{u}_{\max }}\\   
 && u\left( t \right)\in \mathcal{U}:=\left\{ u\in {{L}_{\infty }}:u\left( t \right)\in \Omega \subset {{\mathbb{R}}} \right\}\\
  && \Omega=[{{u}_{\min }},{{u}_{\max}}]\\   
\end{aligned}
\end{equation*}
where $\Gamma$ and $\eta$ in the performance index (cost functional) $J$ are non-negative coefficients, and ${{t}_{f}}$ is the free final time to be optimized. The second term of $J$ is a common choice for the cost functional in molecular control, which measures the energy of the control field in the interval $[t_0,t_f]$. Moreover, the control constraint is represented by a sufficiently smooth function with two–sided interval bounds.

First, let turn our attention to transform the defined $OCP_c$ into an unconstrained optimal control problem $OCP_u$. Following the procedure that originally has been explained in \cite{graichen2010handling}, and has also been used in \cite{d2021pontryagin}, we define a new unconstrained control variable $\nu$ by replacing the control constraint with a smooth and monotonically increasing saturation function as $u=\phi(\nu)$ where
\begin{equation}\label{17}
    {{\phi }}\left( {{\nu}} \right)=u_{max}-\frac{u_{max}-u_{min}}{1+\exp \left( s{{\nu}} \right)}\quad\textit{with}\quad s=\frac{c}{u_{max}-u_{min}}
\end{equation}
in which $c$ is a constant parameter. Since the control variable is changed, we need to consider the following 4 items to define the new unconstrained optimal control problem;
\begin{itemize}
 \item Adding a regularization term to the cost function $J$ defined in $OCP_c$ by considering a regularization parameter $\mu$. 
 
 Therefore the new cost $\Tilde{J}$ is defined as 
     \begin{equation}\label{18}
       \Tilde{J}=J+\mu \int_{{{t}_{0}}}^{{{t}_{f}}}{{{\left\| {{\nu}} \right\|}^{2}}dt} 
     \end{equation}
 As it is explained in details in the optimization by indirect methods, \cite{antony2018large}, the closer the value of $\mu$ gets to zero, the more the solution of the new $OCP_u$ approaches the solution of the original $OCP_c$.      
 \item Introducing an additional multiplier $\varphi$ for taking the equality constraints into account. 
  Hence, the new Pontryagin Hamiltonian is 
  \begin{equation}\label{19}
   \Tilde{ \mathcal{H}}={ \mathcal{H}}+\mu{{\left\| {{\nu }} \right\|}^{2}}+{\varphi\left(u-{{\phi }}\left( \nu \right) \right)}
   \end{equation}    
 \item Considering an additional optimality condition for the new control variable by maximizing new Pontryagin Hamiltonian with respect to $\nu$.
\item Adding the constraint equation
      \begin{equation}\label{21}
        u-\phi(\nu)=0
      \end{equation}
      to the boundary value problem.  
\end{itemize}
Now, the new $OCP_u$ is well-defined. Therefore, the new cost functional is expressed by
\begin{equation}
    \Tilde{J} = \Gamma\, {{t}_{f}}+\eta\int_{{{t}_{0}}}^{t_f}{{{u}^{2}}\left( t \right)\,dt}+\mu \int_{{{t}_{0}}}^{{{t}_{f}}}{{{\left\| {{\nu}} \right\|}^{2}}dt} 
\end{equation}
and the Pontryagin Hamiltonian is formulated as 
\begin{equation}
\begin{aligned}
&\!\!\!\!\!\!\!\Tilde{\mathcal{H}}(\left| \tilde{\psi }\left( t \right) \right\rangle,u(t),\lambda (t),t)=\\
 & \!\!\!\!\!\!\! \left| \lambda \left( {t} \right) \right\rangle ^{{T}} \tilde{H}\left( u\left( t \right) \right)\left| \tilde{\psi }\left( t \right) \right\rangle+\eta {{u}^{2}}\left( t \right)+\mu{{\nu }^{2}}+{\varphi\left(u-{{\phi }}\left( \nu \right) \right)}\\
\end{aligned}
\end{equation}
We now apply the maximum principle to determine the form of the optimal control by
\begin{equation}
    \frac{\partial \Tilde{\mathcal{H}}}{\partial u}=\left| \lambda \left( {t} \right) \right\rangle ^{{T}} \tilde{H}_u\left| \tilde{\psi }\left( t \right) \right\rangle+2\eta {{u}}\left( t \right)+\varphi=0
\end{equation}
where ${{\tilde{{H}}}_{u}}=\frac{\partial \tilde{{H}}\left( u\left( t \right) \right)}{\partial u\left( t \right)}$, and also we have
\begin{equation}
    \frac{\partial \Tilde{\mathcal{H}}}{\partial \nu}=2\mu\nu-\varphi{{\phi }^\prime}\left( \nu \right)=0
\end{equation}
Moreover, we need to apply the first-order necessary conditions for the state and adjoint, so
\begin{equation}\label{37}
    {\left| \dot { \tilde{\psi }} \right\rangle}=\frac{\partial \tilde{\mathcal{H}}}{\partial \left| \lambda  \right\rangle}=\tilde{H}\left( u\left( t \right) \right) \left| \tilde{\psi }\left( t \right) \right\rangle, \quad \left|  \Tilde{\psi} \left( {{t}_{0}} \right) \right\rangle =\left| {{\psi }_{0}} \right\rangle
\end{equation}
\begin{equation}\label{38}
    \dot{ \left| \lambda  \right\rangle }^T=-\frac{\partial \tilde{\mathcal{H}}}{\partial \left| \tilde{\psi } \right\rangle}=-\left| \lambda \left( {t} \right) \right\rangle ^{{T}} \tilde{H}\left( u\left( t \right) \right)
\end{equation}
since the implemented quantum-mechanical Hamiltonian is skew symmetric, we can rewrite \eqref{38} as
\begin{equation}\label{39}
    \dot{ \left| \lambda  \right\rangle }=\tilde{H}\left( u\left( t \right) \right) \left| \lambda\left( t \right) \right\rangle,\quad \left|  \lambda \left( t_f\right) \right\rangle = \bf{0}
\end{equation}
Notice that since the Lagrangian in the cost functional is not dependent on the state, the adjoint equation satisfies the same dynamics as the system state, and, therefore, ${\left| \dot { \tilde{\psi }} \right\rangle}$ and $\dot{ \left| \lambda  \right\rangle }$ are only coupled by means of the control field. Since the system dynamics is not explicitly dependent on time, the Pontryagin Hamiltonian has to be constant. From the transversality condition $\Tilde{\mathcal{H}}\left( t_f \right)=-\Gamma$. 

The presented optimal control problem is solved through the X-TFC framework. To do so, the state and adjoint are approximated using the constrained expressions, hence,
\begin{equation}\label{40}
\begin{aligned}
  &{\left| { \tilde{\psi }} \right\rangle}\left( z,\xi  \right)={{\left( {{h}_{\psi}}\left( z \right)-{{\Omega}_{1}}\left( z \right){{h}_{\psi}}\left( {{z}_{0}} \right)-{{\Omega }_{2}}\left( z \right){{h}_{\psi}}\left( {{z}_{f}} \right) \right)}^{T}}{{\xi }_{\psi}}\\
  &\quad+{{\Omega}_{1}}\left( z \right){\left| {{\psi }_{0}} \right\rangle}+{{\Omega}_{2}}\left( z \right){\left| {{\psi }_{f}} \right\rangle} \\ 
\end{aligned}
\end{equation}
\begin{equation}\label{41}
\begin{aligned}
  \left| \lambda \right\rangle \left( z,\xi \right)= {{\left( {{h}_{\lambda}}\left( z \right)-{{\Omega }_{2}}\left( z \right){{h}_{\lambda}}\left( {{z}_{f}} \right) \right)}^{T}}{{\xi }_{\lambda}}+{{\Omega }_{2}}\left( z \right){\left| {{\lambda}_{f}} \right\rangle}
\end{aligned}
\end{equation}
where $\Omega_1$ and $\Omega_2$ are switching functions, \cite{johnston2020fuel}, which in the general domain of z for initial and final values are respectively expressed as the following:
\begin{equation}\label{42}
 {{\Omega }_{1}}\left( z \right)=1+\frac{2{{\left( z-{{z}_{0}} \right)}^{3}}}{{{\left( {{z}_{f}}-{{z}_{0}} \right)}^{3}}}-\frac{3{{\left( z-{{z}_{0}} \right)}^{2}}}{{{\left( {{z}_{f}}-{{z}_{0}} \right)}^{2}}}   
\end{equation}
\begin{equation}\label{43}
{{\Omega }_{2}}\left( z \right)=-\frac{2{{\left( z-{{z}_{0}} \right)}^{3}}}{{{\left( {{z}_{f}}-{{z}_{0}} \right)}^{3}}}+\frac{3{{\left( z-{{z}_{0}} \right)}^{2}}}{{{\left( {{z}_{f}}-{{z}_{0}} \right)}^{2}}} 
\end{equation}
The derivatives of \eqref{40} and \eqref{41} are, then, 
\begin{equation}\label{}
\begin{aligned}
  &{\left| \dot { \tilde{\psi }} \right\rangle}\left( z,\xi  \right) = c{{\left( {{h}^{\prime}_{\psi}}\left( z \right)-{{\Omega }_{1}^{\prime}}\left( z \right){{h}_{\psi}}\left( {{z}_{0}} \right)-{{\Omega }_{2}^{\prime}}\left( z \right){{h}_{\psi}}\left( {{z}_{f}} \right) \right)}^{T}}{{\xi }_{\psi}}\\
  &\quad+{{\Omega }_{1}^{\prime}}\left( z \right){\left| {{\psi }_{0}} \right\rangle}+{{\Omega }_{2}^{\prime}}\left( z \right){\left| {{\psi }_{f}} \right\rangle} \\ 
\end{aligned}
\end{equation}
\begin{equation}\label{}
\begin{aligned}
 \dot{ \left| {{\lambda }} \right\rangle}  \left( z,\xi \right) = c{{\left( {{h}^{\prime}_{\lambda}}\left( z \right)-{{\Omega }_{2}^{\prime}}\left( z \right){{h}_{\lambda}}\left( {{z}_{f}} \right) \right)}^{T}}{{\xi }_{\lambda}}+{{\Omega}_{2}^{\prime}}\left( z \right){\left| {{\lambda}_{f}} \right\rangle}
\end{aligned}
\end{equation}
Regarding the control variables, the following constrained expressions can be introduced
\begin{equation}\label{44}
 u\left( z,\xi  \right)=h_{u}^{T}\left( z \right){{\xi }_{u}} \\ 
\end{equation}
\begin{equation}\label{45}
 \nu\left( z,\xi \right)=h_{\nu}^{T}\left( z \right){{\xi }_{\nu}} \\ 
\end{equation}
and the multiplier $\varphi$ is expanded as 
\begin{equation}\label{46}
 \varphi\left( z,\xi \right)=h_{\varphi}^{T}\left( z \right){{\xi }_{\varphi}} \\ 
\end{equation}
Actually, the consideration for $\varphi$ can also be done for the control variable $\nu$.
Hence, the vector of PoNN’s unknown parameters to be learned is made up of 
\begin{equation}
    \xi =\left\{ \begin{matrix}
   {{\xi }_{\psi }} & {{\xi }_{\lambda }} & {{\xi }_{u}} & {{\xi }_{\upsilon }} & {{\xi }_{\varphi }} & {c}  \\
\end{matrix} \right\}^T
\end{equation}
and the loss functions to be minimized are 
\begin{equation}
 \mathcal{L}_\psi={\left| \dot { \tilde{\psi }} \right\rangle}-\tilde{H}\left( u\left( t \right) \right) \left| \tilde{\psi }\left( t \right) \right\rangle
\end{equation}
\begin{equation}
\mathcal{L}_\lambda=\dot{ \left| \lambda  \right\rangle }-\tilde{H}\left( u\left( t \right) \right) \left| \lambda\left( t \right) \right\rangle  
\end{equation}
\begin{equation}
\mathcal{L}_u=\left| \lambda \left( {t} \right) \right\rangle ^{{T}} \tilde{H}_u\left| \tilde{\psi }\left( t \right) \right\rangle+2\eta {{u}}\left( t \right)+\varphi
\end{equation}
\begin{equation}
\mathcal{L}_\nu=2\mu\nu-\varphi{{\phi }^\prime}\left( \nu \right)
\end{equation}
\begin{equation}
\mathcal{L}_\phi=u-\phi(\nu)
\end{equation}
and the loss function being computed at the final time is
\begin{equation}
\mathcal{L}_{\Tilde{\mathcal{H}}}=\Tilde{\mathcal{H}}\left( t_f \right)+\Gamma
\end{equation}
Now, we have to impose the valid solution of the augmented loss function as a $N \times \Tilde{N}$ dimensional zero matrix, and solve it by means of a numerical minimization scheme.
\section{Simulation Results}
\begin{figure}[thpb]
  \centering
  \includegraphics[scale=0.6]{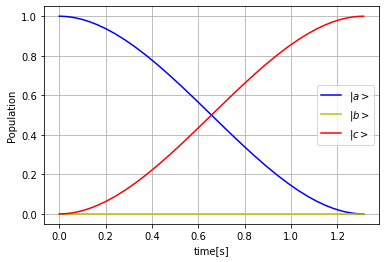}
  \caption{Population dynamics under quantum dynamics}
  \label{population}
\end{figure}

\begin{figure}[thpb]
  \centering
  \includegraphics[scale=1]{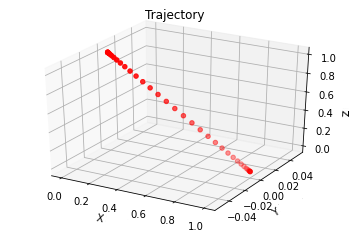}
  \caption{State trajectories in Cartesian coordinates}
  \label{trajectory}
\end{figure}

\begin{figure}[thpb]
  \centering
  \includegraphics[scale=0.6]{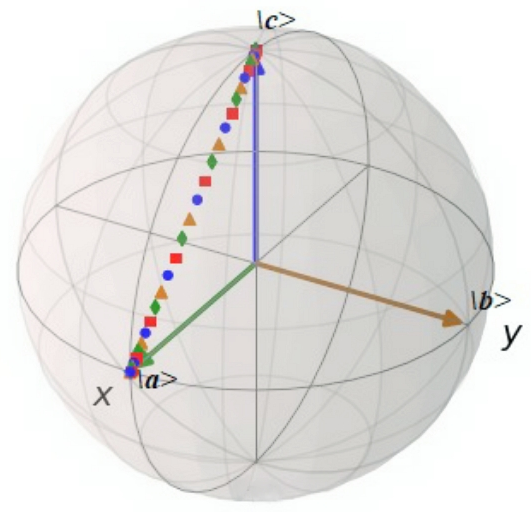}
  \caption{Quantum state evolution within spherical coordinates}
  \label{sphere}
\end{figure}

\begin{figure}[thpb]
  \centering
  \includegraphics[scale=0.6]{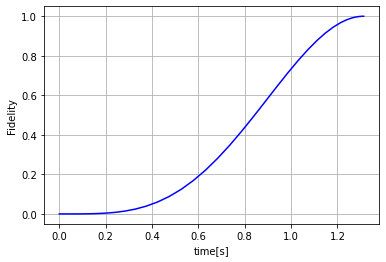}
  \caption{Quantum state transition probability}
  \label{fidelity}
\end{figure}

We consider a three-level system being controlled through a single control field $u(t)$ 
on the orthonormal basis $\left| a \right\rangle ={{\left[ \begin{matrix}
  1 & 0 & 0  \\ \end{matrix} \right]}^{T}}$, $\left| b \right\rangle ={{\left[ \begin{matrix}
  0 & 1 & 0  \\ \end{matrix} \right]}^{T}}$, and $\left| c \right\rangle ={{\left[ \begin{matrix}
  0 & 0 & 1  \\ \end{matrix} \right]}^{T}}$.
%\begin{equation}
%  \left\{ \begin{matrix}
% \left| a \right\rangle ={{\left[ \begin{matrix}
%   1 & 0 & 0  \\
%\end{matrix} \right]}^{T}}  \\
%   \left| b \right\rangle ={{\left[ \begin{matrix}
%   0 & 1 & 0  \\
%\end{matrix} \right]}^{T}}  \\
%   \left| c \right\rangle ={{\left[ \begin{matrix}
%   0 & 0 & 1  \\
%\end{matrix} \right]}^{T}}  \\
%\end{matrix} \right. 
%\end{equation}
The drift and control Hamiltonians for such system are expressed as, \cite{cong2014control},
\begin{equation}\label{}
 {{H}_{d}}=\left( \begin{matrix}
   0.5 & 0 & 0  \\
   0 & 0.4 & 0  \\
   0 & 0 & 0.6  \\
\end{matrix} \right), \quad {{H}_{C}}\left( t \right)=u\left( t \right)\left( \begin{matrix}
   0 & 1 & 0  \\
   1 & 0 & 1  \\
   0 & 1 & 0  \\
\end{matrix} \right)
\end{equation}
The convergence tolerance is considered on $10^{-16}$, $N=40$, $\eta=0.1$, and $\Gamma=0.01$. Moreover we use the Chebyshev orthogonal polynomials as the basis for tfc implementation. However, different basis or number of discretization points can be experimented. Let consider the transition of the initial state $\left| {{\psi }_{i}} \right\rangle=\left|a\right\rangle$ to the target $\left| {{\psi }_{d}} \right\rangle=\left|c\right\rangle$, for the case that $-1\le u(t) \le 1 $. The population evolution is shown in Fig \ref{population}. As it can be seen from the graph, the population of the initial state decays exponentially, while the population of target increases by time.
Figure \ref{trajectory} shows the state trajectories in Cartesian coordinates, which is learnt by the PoNN during the computation time.
To better understand the trajectory, we also considered the quantum state evolution on the Bloch sphere with the spherical coordinates in Fig \ref{sphere}.
Finally, in order to check the security level of reaching the target, we have to calculate the transition probability known as the quantum fidelity, which for two pure quantum states is computed by, \cite{liang2019quantum},
\begin{equation}
 {\mathcal{F}(\left| {\psi} \left( t \right) \right\rangle,\left| {\psi}_f\right\rangle)=|\langle {\psi} (t)|\psi _{f}\rangle |^{2}}  
\end{equation}
Since the system studied in this paper is a closed quantum system, we expect to reach the fidelity of 1, as well shown in Fig \ref{fidelity}.
\section{CONCLUSIONS}
In this paper, we have exploited the newly introduced method called the physics informed theory of functional connections. Through this method, PoNNs are used to learn optimal control actions for an input constrained optimal control problem. We used the scheme to solve the boundary value problem arising from the application of the indirect optimal control method to a quantum state to state transition problem under control constraint at the cost of compromising between the goal of attaining optimal time and, simultaneously, keeping the energy of the field relatively small. For the studied quantum control system, we considered a pure state, evolving through Schrödinger equation. The same procedure can also be applied for considering the dynamics of the evolution operator. Future challenges first consist in a detailed comparison and illustration the superiority of PoNN method over the state of the art quantum control methods, and also exploiting the versatility of the optimal control paradigm further by considering an open quantum system, for which the dynamics is explained through, e.g., Markovian or stochastic master equations, and also considering additional constraints, e.g., state constraints, to tackle with more complicated quantum optimal control problems.

\addtolength{\textheight}{-12cm} 

%%%%%%%%%%%%%%%%%%%%%%%%%%%%%%%%%%%%%%%%%%%%%%%%%%%%%%%%%%%%%%%%%%%%%%%%%%%%%%%%

\section*{ACKNOWLEDGMENT}
The authors acknowledge the support of FCT for the grant 2021.07608.BD, the ARISE Associated Laboratory, Ref. LA/P/0112/2020, and the R$\&$D Unit SYSTEC-Base, Ref. UIDB/00147/2020, and Programmatic, Ref. UIDP/00147/2020 funds, and also the support of projects SNAP, Ref. NORTE-01-0145-FEDER-000085, and  RELIABLE (PTDC/EEI-AUT/3522/2020) funded by national funds through FCT/MCTES. The work has been done in the honor and memory of Professor Fernando Lobo Pereira.

\end{document}